\documentclass[aps,pra,amsmath,amssymb,tightenlines,epsfig,floatfix,superscriptaddress, longbibliography]{revtex4}

\usepackage{graphicx}
\usepackage{dcolumn}
\usepackage{bm}
\usepackage{physics}
\usepackage{amsfonts}
\usepackage{xspace}
\usepackage{color}
\usepackage{xcolor}
\usepackage{epstopdf}
\usepackage{multirow}
\usepackage{hyperref}

\usepackage{upgreek}

\begin{document}
\newcommand{\sah}[1]{\textcolor{red}{#1}}

\title{An atomic Fabry-Perot interferometer-based acceleration sensor for microgravity environments}

\author{Manju Perumbil}
\email{manjuperumbil@cukerala.ac.in}
\affiliation{ITEP, Department of Education, Central University of Kerala, \\Tejaswini Hills, Periye, Kasaragod, Kerala, India.}
\affiliation{Department of Quantum Science, Australian National University, Canberra, Australia, 2602.}
 
\author{Matthew J. Blacker}%
 \affiliation{Department of Quantum Science, Australian National University, Canberra, Australia, 2602.}
\affiliation{Department of Applied Mathematics and Theoretical Physics, University of Cambridge, Cambridge, CB3 0WA, UK.}

\author{Stuart S. Szigetti}
\affiliation{Department of Quantum Science, Australian National University, Canberra, Australia, 2602.}%

\author{Simon A. Haine}
\affiliation{Department of Quantum Science, Australian National University, Canberra, Australia, 2602.}%

\date{\today}

\begin{abstract}
We investigate the use of an atomic Fabry-Perot interferometer (FPI) with a pulsed non-interacting Bose-Einstein condensate (BEC) source as a space-based acceleration sensor. We derive an analytic approximation for the device's transmission under a uniform acceleration, which we use to compute the device's attainable acceleration sensitivity using the classical Fisher information. In the ideal case of a high-finesse FPI and an infinitely narrow momentum width atomic source, we find that when the total length of the device is constrained to small values, the atomic FPI can achieve greater acceleration sensitivity than a Mach-Zender (MZ) interferometer of equivalent total device length. Under the more realistic case of a finite momentum width atomic source, We identify the ideal cavity length that gives the best sensitivity. Although the MZ interferometer now offers enhanced sensitivity within currently-achievable experimental parameter regimes, our analysis demonstrates that the atomic FPI holds potential as a promising alternative in the future, provided that narrow momentum width atomic sources can be engineered.
\end{abstract}


\maketitle

\section{Introduction}\label{sec1}
The existing generation of atom interferometers have provided state-of-the-art measurements of accelerations~\cite{Canuel:2006, Templier:2022}, rotations~\cite{Gustavson:1997,Gustavson:2000,Durfee:2006,Gauguet:2009,Gautier:2022}, gravitational fields~\cite{peters_measurement_1999,peters_high-precision_2001,altin_precision_2013, Hu:2013,Farah:2014, hardman_simultaneous_2016,Zhang:2023b}, gravity gradients~\cite{Snadden:1998, sorrentino_sensitivity_2014,Biedermann:2015,damico_bragg_2016,Asenbaum:2017,Janvier:2022}, the fine structure constant~\cite{parker_measurement_2018,yu_atom-interferometry_2019,morel_determination_2020}, and Newton's gravitational constant~\cite{Rosi:2014}. With sufficient miniaturization and ruggedization, quantum sensors based on atom interferometry could enable new capabilities in navigation~\cite{Jekeli:2005, Battelier:2016,Narducci:2022, wang_enhancing_2021, Wright:2022, Phillips:2022}, civil engineering risk management \cite{metje_seeing_2011,boddice_capability_2017,Stray:2022}, mineral exploration and recovery~\cite{van_leeuwen_bhp_2000,bongs_taking_2019}, groundwater mapping and monitoring~\cite{schilling_gravity_2020}, and geodesy~\cite{Stockton2011, Migliaccio:2019, Trimeche:2019,Leveque:2021}. Atom interferometers are presently being developed for mobile operation on dynamic platforms, and have been deployed on ships~\cite{Bidel:2018,Wu:2023}, aircraft~\cite{Geiger2011, Bidel:2020, Bidel:2023}, and in microgravity environments onboard sounding rockets~\cite{becker_space-borne_2018,Lachmann:2021} and the International Space Station~\cite{williams_pathfinder_2024}. Spaceborne operation in particular has provided a strong motivation for next-generation atom-interferometric development, since it could progress key questions in fundamental physics through low-frequency-band gravitational wave detection~\cite{dimopoulos_atomic_2008}, weak equivalence principle violation tests~\cite{dimopoulos_testing_2007, williams_quantum_2016}, and novel experiments into dark energy~\cite{burrage_using_2016,sabulsky_experiment_2019}, dark matter~\cite{geraci_sensitivity_2016,badurina_aion_2020}, and quantum gravity~\cite{Haine:2021,margalit_realization_2021}.

There is a worldwide effort to decrease the size, weight, and power (SWaP) of atomic inertial sensors whilst maintaining sensitivity, accuracy, and stability on dynamic platforms in real-world environments~\cite{fang_metrology_2016,bongs_taking_2019, Geiger:2020, Narducci:2022}. Efforts to address these challenges have largely focussed on improving the performance of the standard three-pulse Mach-Zehnder (MZ) atom interferometer, through innovations such as large momentum transfer atom optics~\cite{Clade:2009, Muller:2008b, Chiow:2011, McDonald:2013b, Kotru:2015, Gebbe:2021, Wilkason:2022, Beguin:2023}, improved atomic source quality and production rate~\cite{Debs:2011, robins_atom_2013, szigeti_why_2012, Deppner:2021, Hensel:2021,Chen:2022, Lee:2022}, novel state readout~\cite{wigley_readout-delay-free_2019, Piccon:2022, benaicha:2024}, error-robust quantum control~\cite{Saywell:2020, Saywell:2023, Saywell:2023a, Wang:2024, Rodzinka:2024}, and overcoming the shot-noise limit through quantum entanglement generated from atom-atom \cite{esteve_squeezing_2008,appel_mesoscopic_2009,lucke_twin_2011, Haine:2011, hamley_spin-nematic_2012,lucke_detecting_2014,muessel_twist-and-turn_2015,lange_entanglement_2018, Szigeti:2020, Szigeti:2021} or atom-light~\cite{hald_spin_1999,leroux_implementation_2010,schleier-smith_squeezing_2010,sewell_magnetic_2012, Haine:2013, Szigeti:2014b, Haine:2016, Kritsotakis:2021, Fuderer:2023} interactions. However, another approach is to consider alternatives to the standard MZ atom interferometer, which could relax certain technological requirements and provide advantages in tight-SWaP situations.

One alternative interferometry configuration is the atomic analogue of a Fabry-Perot interferometer (FPI). In an optical FPI, light enters a cavity formed by two parallel mirrors, and a resonant spectra is obtained by scanning the incident wavelength. Optical FPIs have been extensively used in a number of spectroscopic \cite{drever_laser_1983,deventer_comparison_1990,xue_pulsed_2016} and sensing \cite{yoshino_fiber-optic_1982,taylor_principles_1998} applications. In an atomic FPI, the incoming light is replaced by atomic matter-waves and the mirrors are replaced by laser-induced potential barriers. Previous theoretical studies into atomic FPIs have demonstrated that an ultracold Bose source can produce high contrast Fabry-Perot interference fringes~\cite{carusotto_nonlinear_2001, paul_nonlinear_2005, paul_nonlinear_2007, rapedius_barrier_2008, ernst_transport_2010}, characterised their resonance properties \cite{dutt_smooth_2010,damon_reduction_2014}, and investigated the potential use of atomic FPIs in velocity selection~\cite{wilkens_Fabry-Perot_1993,ruschhaupt_velocity_2005} and angle selection~\cite{valagiannopoulos_quantum_2019}. For the experimental regimes achievable with current technology, a narrow momentum width source such as a Bose-Einstein condensate (BEC) is needed to achieve the high contrast resonant transmission peaks required for useful sensing~\cite{manju_atomic_2020}. Operating in a regime where atom-atom collisional interactions are negligible is also highly desirable, and can be obtained through a Feshbach resonance~\cite{Roberts:1998,Kuhn:2014,Everitt:2017}. The suitability of a noninteracting BEC source for atomic Fabry-Perot interferometry has been validated in a recent experimental demonstration using a $^{39}$K BEC source and optical barrier potentials formed using a digital micromirror device~\cite{Eid:2024}.

\begin{figure}[!t!!]
\includegraphics[width=0.7\columnwidth]{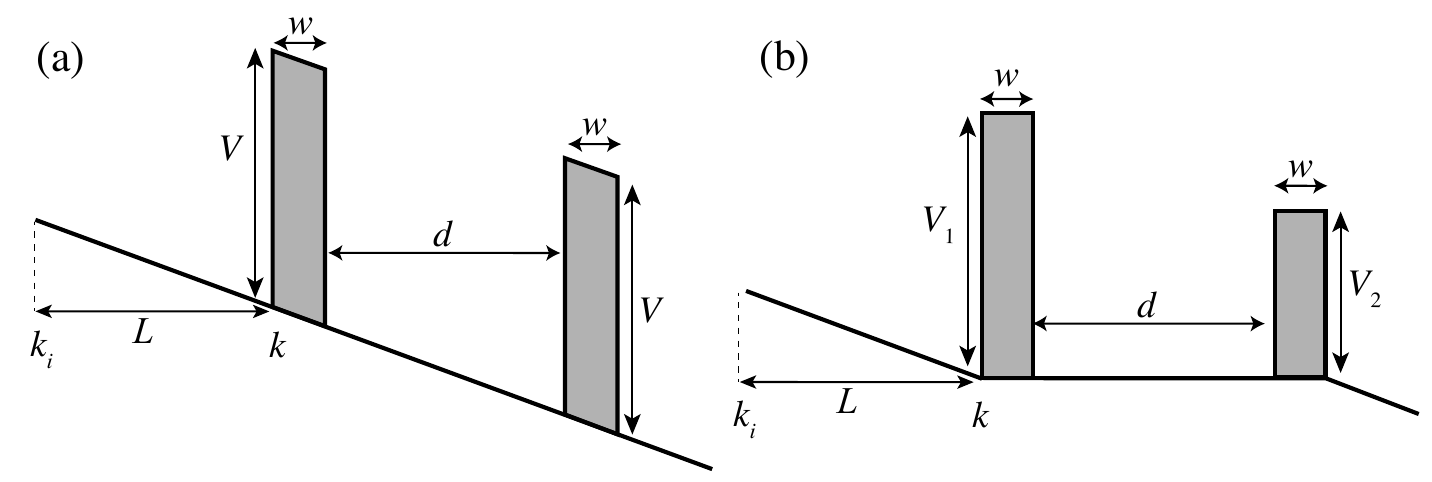}
	\caption{\label{fig:SchematicFPIAccelerating}(a) Schematic diagram of an atomic FPI made of two symmetric rectangular barriers in an accelerating field. (b) Simplified model, with an exact analytic solution, that approximates the system shown in (a). Here $\hbar k_i$ is the initial momentum of the particle, $\hbar k$ is the momentum of the particle at the position of the first barrier, after travelling a distance $L$, $V_1$ and $V_2$ are the heights of the first and second barriers respectively, $w$ is the barrier width, and $d$ is the cavity length.}	
\end{figure} 

In this paper, we study the application of an atomic FPI as an acceleration sensor.  A previous work has considered the suitability of atomic FPIs for gravimetry through a numerical simulation analysis~\cite{schach_tunneling_2022}. Here we take a complementary analytic approach that aims to give deeper insights into the optimal parameter regime and performance limits of an atomic FPI acceleration sensor. This approach allows us to assess whether there are regimes where an atomic FPI could potentially offer superior performance as an acceleration sensor compared to MZ interferometry. We are particularly interested in situations where device size is highly constrained, such as in space-based applications. Since we are interested in fundamental performance limits, we consider only the case of a non-interacting BEC in this work.

Specifically, in Section \ref{sec:methods} we derive an analytic expression for the transmission of a non-interacting BEC through the atomic FPI in an accelerating field, which we validate by numeric simulation of the Schr\"{o}dinger equation for $N$ non-interacting particles. We first study in Section \ref{Sec31} the ideal case of an infinitely narrow momentum width source, and from the transmission derive an approximate expression for the optimum Fisher information (and consequently acceleration sensitivity). We then consider in Sections \ref{sec:finitemomentumwidth} and \ref{sec:Optdfordeltak} atomic clouds with a finite momentum width, and study the effect of finite momentum width upon the free parameters which lead to optimal Fisher information. In each case, in Sections \ref{Sec32} and \ref{sec:finitewidthcomparison} respectively we compare the acceleration sensitivity of a space-based atomic FPI to a space-based MZ interferometer of equivalent device size, to assess the future potential of an atomic FPI as an alternative accelerometry device.\\

\section{Methods}
\label{sec:methods}
\subsection{Model} \label{sec:model}
We consider the transmission of a beam of particles through an atomic Fabry-Perot `cavity' made of two symmetric rectangular barriers in a uniform accelerating field (see Fig.~\ref{fig:SchematicFPIAccelerating}(a)).
We obtain a simplified model that is analytically tractable by making two assumptions. Firstly, we assume that in each barrier the linear variation in the acceleration can be neglected, and also the acceleration experienced by the particles within each barrier is negligible. This assumption requires the barrier width ($w$) to be small compared to the distance between the initial particle positions and the first barrier ($L$). Secondly, we account for the acceleration potential energy the particles gain after travelling a distance $w + d$ through the first barrier and the cavity by reducing the energy of the second barrier by the particle's energy gain:
\begin{align}
    V_2(a)=V_1-ma(w+d), \label{V2Eq}
\end{align}
where $m$ is the mass of each particle and $a$ is acceleration 
 (assumed to be uniform over device length $d + 2w$), $V_1$ and $V_2$ are the heights of the first and second barriers, respectively. These two simplifications give the double asymmetric rectangular barrier model shown in Fig.~\ref{fig:SchematicFPIAccelerating}(b).\\

Using this simplified model, we can analytically determine how acceleration affects transmission through the atomic FPI. We first consider the case of an incoming plane wave. If the initial momentum of the particle is $\hbar k_i$, then classically after travelling a distance $L$ under uniform acceleration $a$, the particle's energy changes by $m a L$. Since the BEC will be prepared a distance $L$ from the first barrier, we therefore take the momentum of the plane wave incident on the asymmetric double barrier system to be
\begin{align}
    \hbar k(a) = \hbar k_i \sqrt{1 + \frac{2m^2 a L}{\hbar^2 k_i^2}}.
    \label{eq:kainki}
\end{align}
The probability of transmission through the double barrier system is given by the transmission coefficient~\cite{xiao_resonant_2015}
\begin{align}
    T_{k_i}(a) = \frac{T_{\text{max}}(a)}{1 + \left( \frac{2 \mathcal{F}(a)}{\pi} \right)^2 \sin^2 \left( k(a) d + \phi_a(a) \right)},
\label{eq:Transmissioncoefficientk}
\end{align}
where $T_{\text{max}}(a)$ is the maximum achievable transmission coefficient, $\mathcal{F}(a)$ is the finesse of the atomic Fabry-Perot cavity and $\phi_a(a)$ is a phase shift that sets the resonance condition for the cavity. Our decision to denote the dependence of $k_i$ explicitly will become clear shortly. In analogy with the optical Fabry-Perot cavity, the maximum transmission coefficient and finesse are most intuitively expressed in terms of the reflection coefficients of the two barriers (i.e. cavity `mirrors'), $R_1(a)$ and $R_2(a)$. Explicitly,
\begin{align}
    T_{\text{max}}(a)   &= 1 - \left[ \frac{\sqrt{R_1(a)} - \sqrt{R_2(a)}}{1 - \sqrt{R_1(a) R_2(a)}} \right]^2, \\
    \mathcal{F}(a)  &= \frac{\pi \left[ R_1(a) R_2(a) \right]^{1/4}}{1 - \sqrt{R_1(a) R_2(a)}},
\end{align}
where
\begin{align}
    R_j(a) = \frac{{M_{j}^+(a)}^2}{{M_{j}^-(a)}^2 + \coth^2\left[\beta_j(a) w \right]},
\end{align}
with
\begin{subequations}
\begin{align}
   \beta_1(a)^2=\frac{2m}{\hbar^2}[V_1-E(a)], \quad \beta_2(a)^2=\frac{2m}{\hbar^2}[V_2(a)-E(a)], 
  \end{align}
 \begin{align}
  M_{j}^{\pm}(a) = & \frac{1}{2} \left[ \frac{\beta_j(a)}{k(a)} \pm \frac{k(a)}{\beta_j(a)} \right].    
 \end{align} 
\end{subequations} 
Here $E(a) = \left( \hbar k(a) \right)^2/2m$ is the energy of the incident plane wave. The phase shift $\phi_a(a)$ similarly depends upon the cavity `mirror' parameters:
\begin{align}
    \phi_a(a) = \frac{1}{2} \left[ \pi - \left( \phi_1(a) + \phi_2(a) \right) \right],
\end{align}
where
\begin{align}
    \phi_j(a) = \tan^{-1} \left[ M_j^{-} \tanh \left( \beta_j(a) w_j \right) \right], j=1, 2.
\end{align}

\begin{figure}[t!]
	\includegraphics[width=1\linewidth]{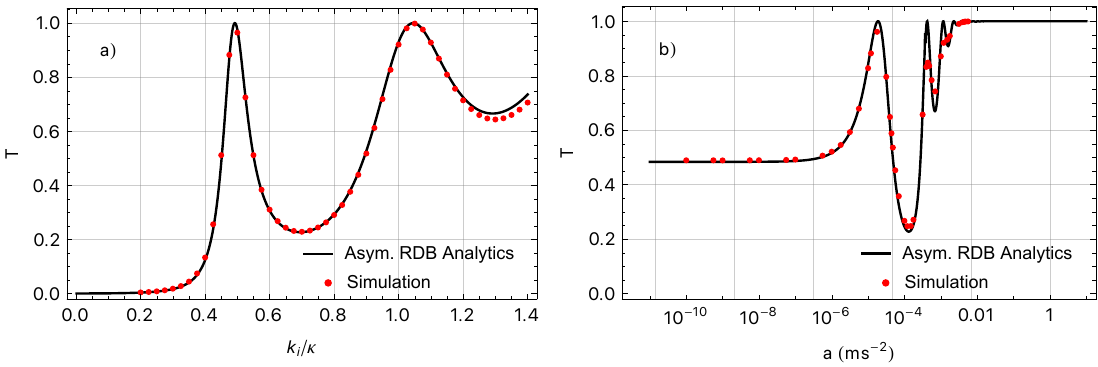}
\caption[Comparison of the analytic and simulation results for the transmission spectra of an atomic FPI in an accelerating field]{Comparison of the analytic (black curve) and simulation (red dotted curve) results of transmission resonances of a beam of particles transmitting through double rectangular barriers in an accelerating field, for a) a fixed acceleration $a=0$ and varying initial momentum kick $k_i/\kappa$ and b) a fixed initial momentum kick $k_i/\kappa = 0.45$ and varying acceleration $a$. Here, the Schr\"{o}dinger equation is simulated for a Gaussian cloud of a non-interacting BEC source. The parameters used for the simulation are: $V_1=3.83 \times 10^{-32} J$, $w=1 \upmu$m and $d=4 \upmu$m. For $^{85}$Rb, the momentum scale $\kappa=\sqrt{2 m V_1}/\hbar =9.9 \times 10^6 \text{m}^{-1}$. The external potential is given by $V(z)=V_b-maz$, where $V_b$ is the double rectangular barrier potential. The analytical results agree well with the simulation result, validating the analytical model. } 
	\label{fig:TvskAcceleratedRDB}
\end{figure} 

The atomic system provides key differences compared to the well-known optical FPI, including atomic mass and mirrors where the key parameters can be tuned. Hence the transmission spectrum of an atomic FPI differs from that of an optical FPI in many ways. The reflectivity of optical potentials exhibits distinct behaviour compared to conventional mirrors, resulting in variations in reflectivity when scanning the momentum/energy of the source atoms. Hence, unlike the optical case, in the atomic analog scanning $k$ and $d$ are not equivalent. The width and contrast of the peaks in the transmission spectrum changes with change in wave number of the atomic source. Nevertheless, as we show below, in appropriately chosen regimes the transmission depends sensitively on the acceleration, allowing an atomic FPI to operate as a sensitive accelerometer.

It is straightforward to extend the above results to an incident atomic cloud of non-interacting atoms with a spread of momenta. Since the atoms do not interact, there is a one-to-one mapping between incoming and outgoing momentum. Consequently, the overall transmission is simply the integral over all transmission coefficients (indexed by incident wavevector $k_i$) weighted by the incident atomic cloud's $k$-space distribution $P(k_i)$:
\begin{align}
    T(a)   &= \int dk_i \ P(k_i) T_{k_i}(a).
\end{align}
We validate our analytic model by comparing to numeric simulation. In particular, we simulate the evolution of the Schr\"{o}dinger equation for a non-interacting Gaussian BEC source, which we assume was in the ground state of a harmonic trapping potential of trapping frequency $\omega_z$, before being released and interacting with the atomic Fabrey-Perot cavity, formed by external potential $V(z)$. The transmission coefficient $T$ is computed as
\begin{align}
    T = \frac{N_T}{N_T + N_R},
\end{align}
where $N_T$ and $N_R$, the number of transmitted and reflected particles, are defined as
\begin{subequations}
\begin{align}
 N_T & = \int_{z_T}^{\infty} \abs{\psi(z,t_{\text{end}})}^2 \, dz, \\
 N_R & = \int_{-\infty}^{z_R} \abs{\psi (z,t_{\text{end}})}^2 \, dz. 
\end{align}
\end{subequations}
The transmitted and reflected regions are specified as $z> z_T = z_0 + 3\sigma_c + L + d + 2w$ and $z < z_R = z_0 + 3 \sigma_c + L$ respectively, where $z_0$ denotes the initial position of the atomic cloud. The stopping time, $t_{\text{end}}$ is chosen so that there are no atoms left in the cavity; this is quantified by when $N_T/N$ and $N_R/N$ do not change by more than $10^{-6}$ in a given time step. Here, $N(t) = \int_{-\infty}^{\infty} dz \,\abs{\psi(z,t)}^2$ is the normalization of the wavefunction. The external potential used is given by $V(z) = V_b - maz$, where $V_b$ is the potential generated by two barriers of width $w$ and height $V_1$ separated by distance $d$. The simulation was completed using the open-source software package XMDS2~\cite{dennis_xmds2_2013} with an adaptive 4th-5th order Runge-Kutta interaction picture algorithm.

In Figure \ref{fig:TvskAcceleratedRDB}a) we plot the transmission coefficient calculated via equation \eqref{eq:Transmissioncoefficientk} (black solid curve) and numeric simulation (red dotted curve) corresponding to a range of $k_i/\kappa$ values, where $\kappa$ is a momentum length scale determined by $\kappa = \sqrt{2mV_1/\hbar^2}$. We use the cavity parameters for $^{85}$Rb determined in Ref.~\cite{manju_atomic_2020} ($V_1 = 3.83 \times 10^{-32}$J = $5.81 \hbar \omega_z$, $w=1 \upmu$m and $d=4 \upmu$m) with trapping frequency $\omega_z = 2\pi \times 10 \text{ Hz}$ in the presence of an acceleration of $a = 0$.  Similarly, in Figure \ref{fig:TvskAcceleratedRDB}b) we plot the transmission coefficient for a fixed $k_i/\kappa = 0.45$ and varying $a$. In both instances, we observe that the curve corresponding to equation \eqref{eq:Transmissioncoefficientk} matches the simulation data very well, validating the analytic model.

\subsection{Quantifying Acceleration Sensitivity}
\label{sec:Accelerationsensitivity}
The smallest change in acceleration ($\delta a)$ detectable by an accelerometer quantifies the sensitivity of the device. For a cloud of $N$ non-interacting, uncorrelated atoms, this is given by the Cram\'{e}r-Rao bound \cite{toth_quantum_2014} 
\begin{align}
    \delta a    &=\frac{1}{\sqrt{N F_C(a)}},
\label{eq:deltaa}
\end{align}
where $F_C(a)$ is the per-particle classical Fisher information \cite{haine_mean-field_2016, Kritsotakis:2018}, given by
\begin{equation}
    F_C(a) = \sum_m \frac{(\partial \mathcal{P}_m / \partial a)^2}{\mathcal{P}_m(a)}.
\end{equation}
Here $\mathcal{P}_m(a)$ is the probability distribution (indexed by $m$) constructed from measurements of a particular observable, and so $F_C(a)$ depends upon this choice of observable. For the atomic FPI considered in this work, we measure the number of transmitted and reflected atoms, yielding the transmission and reflection coefficients $T(a)$ and $R(a) = 1 - T(a)$, respectively. These coefficients are the probability distributions for transmission and reflection, respectively, that we need to compute the classical Fisher information:
\begin{align}
\begin{split}
    F_C(a) &=\frac{(\partial T/\partial a)^2}{T(a)}+\frac{(\partial R/\partial a)^2}{R(a)} \\
            &= \frac{(\partial T/\partial a)^2}{T(a)(1-T(a))}, 
\end{split}
\label{eq:FC}
\end{align}
where we have invoked $\partial R/ \partial a = - \partial T/ \partial a$. From Eq.~\eqref{eq:deltaa}, it follows that we should optimize for higher $F_C$, since that corresponds to a more sensitive accelerometer.\\

In our analysis we use dimensionless parameters with $1/\kappa$ as the unit of length and $\hbar \kappa$ as the unit of momentum, where $\kappa=\sqrt{(2mV_1/\hbar)}$ is the wave vector corresponding to the first barrier. Specifically, we define
\begin{align}
     \tilde{d}&=\kappa d, \quad \tilde{L}=\kappa L, \quad\tilde{k}=k/\kappa,   \\
 \tilde{a}&=\frac{2m^2a}{\hbar^2\kappa^3}, \quad \tilde{F}_C=\frac{F_C\hbar^4 \kappa^6}{4m^4}.
\end{align}

\section{Results and Discussion}\label{sec2}
\subsection{Acceleration sensitivity of an atomic FPI with a plane matter-wave input}
\label{Sec31}
We begin our investigation into the sensitivity of an atomic FPI as an accelerometer by considering a BEC source with an infinitely narrow momentum width. This provides intuition for parameter dependencies of the sensitivity in the ideal case which optimises transmission through the FPI \cite{manju_atomic_2020}. $\tilde{F}_C$ varies with $\tilde{k}_i$ and we can estimate optimum $\tilde{k}_i$ that gives the maximum $\tilde{F}_C$ for each cavity length.

\begin{figure}[t!]
	\centering
 \includegraphics[width=0.8\linewidth]{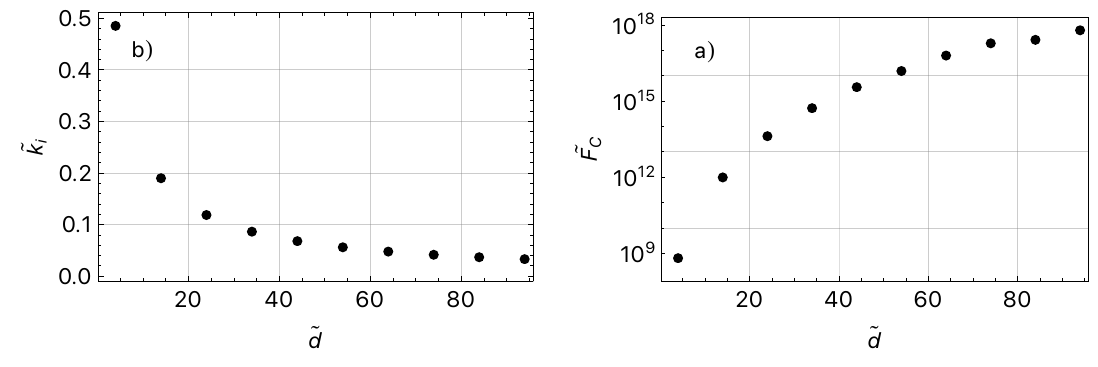}
	\caption[The Fisher information corresponding to the optimum particle momentum and the particle momentum as a function of the cavity length]{Optimum particle momentum (a) and the Fisher information corresponding to the optimum particle momentum (b) as a function of the cavity length, in the case of an infinitely narrow momentum width source of atoms. For each cavity length there exists an optimum $\tilde{k}_i$ and it decreases with increasing cavity length. The optimum value of $\tilde{k}_i$ ranges from 0.06 to 0.004 $k_0$, where $k_0=2\pi/(780\text{nm)}$. The maximum $\tilde{F}_C$ increases with increasing cavity length. Hence in this case, sensitivity to acceleration can be improved by increasing cavity length.   }
	\label{fig:Optk0Fcvsd}
\end{figure}
Fig.~\ref{fig:Optk0Fcvsd} a) shows the variation in optimum $\tilde{k}_i$ as a function of the cavity length (the height of the first barrier is fixed here, resulting in a constant $\kappa$). Here we can see that as the cavity length increases, optimum $\tilde{k}_i$ decreases. This means that, for a fixed barrier height, the optimum momentum of the atoms that gives maximum sensitivity to acceleration decreases with increasing cavity length.
Fig.~\ref{fig:Optk0Fcvsd} b) illustrates the variation in $\tilde{F}_C$ corresponding to optimum $\tilde{k}_i$ as a function of cavity length. This shows that $F_C$ and hence the sensitivity increases with increasing cavity length in the case of a cloud with infinitely narrow momentum width.  \\

The trend in Fig.~\ref{fig:Optk0Fcvsd} arises due to the changes in transmission peak properties with variation in the cavity length. As the cavity length increases, the linewidth of the resonant peaks gets narrower, leading to curves with higher slopes ($\partial T/ \partial k$) \cite{manju_atomic_2020}. The relationship between the slope of the transmission spectra and acceleration sensitivity can be obtained as follows. Under uniform acceleration, the velocity of the cloud at the position of the first barrier is $v=v_0+at$, where $v_0$ is the initial velocity, $t$ is the time taken to reach the first barrier and $a$ is the acceleration. This yields
\begin{align}
 \frac{\partial T}{\partial a}&=\frac{\partial T}{\partial k} \frac{\partial k}{\partial a} = \frac{mt}{\hbar}\frac{\partial T}{\partial k} . 
 \label{Eq.dTda}
\end{align}
Equations \ref{eq:FC} and \ref{Eq.dTda} show that the classical Fisher information increases with increasing $\partial T/ \partial k$. Hence, as the cavity length increases, an increase in the slope ($\partial T/ \partial k$) causes the increase in $F_C$, as observed in Fig.~\ref{fig:Optk0Fcvsd} b). Here, the time $t$ depends on the distance between the initial position of the cloud and the position of the first barrier $L$. Hence, the sensitivity depends on $L$.\\

We now formulate a compact analytic expression for the classical Fisher information that is straightforward to optimize, thereby providing a `best-case' estimate of an atomic FPI accelerometer's sensitivity. For an infinitely narrow source, we obtain the Fisher information by substituting equation \eqref{eq:Transmissioncoefficientk} into equation \eqref{eq:FC}. In calculating $\partial T/ \partial a$, we assume the dependence on acceleration of finesse $\mathcal{F}$ and phase $\phi_a$ is insignificant compared to the dependence on acceleration of $k$. Under that assumption, we obtain 
\begin{align}
\Tilde{F_C}^{\text{approx}}&= \frac{16 \tilde{d}^2 \mathcal{F}^4 \pi^2 T_{\text{max}} \sin^2 \left(2 \Phi(\tilde{a}) \right)\tilde{k}'^2}{\Big( \pi^2(T_{\text{max}}-1)-4\mathcal{F}^2\sin^2 \left( \Phi(\tilde{a}) \right)\Big)(\pi^2+4\mathcal{F}^2\sin^2 \left( \Phi(\tilde{a}) \right))^2},
    \label{eqApproxFC}
\end{align}
where
\begin{subequations}
\begin{align}
   \Phi(\tilde{a}) = & \phi_a + \tilde{k}(\tilde{a}) \tilde{d}, \\
   \tilde{k}' = & \frac{\partial \tilde{k}(\tilde{a})}{\partial \tilde{a}} \, .
 \end{align}
\end{subequations}
In\begin{figure}[t!]
	\centering
	\includegraphics[width=0.5\linewidth]{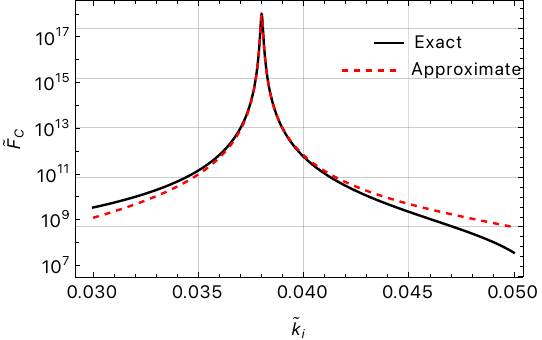}
	\caption[\sah{[Units for $F_c$ again. See comment on previous plot]} Comparison of the exact and approximate expressions of $F_C$]{Comparison of the variation of $\tilde{F}_C$ with $\tilde{k}_i$ using exact (equation \eqref{eq:FC}-dimensionless form) and approximate (equation \eqref{eqApproxFC}) expressions of $\tilde{F}_C$ in the black and red dashed curve respectively. The approximate value of optimum $\tilde{F}_C$ agrees well with the exact value near the optimum region.  Here, $\tilde{d} = 80$ and $\tilde{a} \rightarrow 0$.}
	\label{fig:LogFcApproxFcExactd80New}
\end{figure} 
Figure \ref{fig:LogFcApproxFcExactd80New}, we compare this approximate expression (red dashed curve) to the exact expression (black curve) computed numerically via equation \eqref{eq:FC}. In the region near the optimum (maximum) $\tilde{F}_C$, the curves agree very well, validating the approximate expression in equation \eqref{eqApproxFC}.

We now derive an expression for this optimum $F_C$, and hence the precision to which the acceleration can be inferred. We assume that the acceleration is known approximately ($a\approx a_0$), and we wish to determine small deviations $\delta a$ from this value. That is $a = a_0 + \delta a$. 
%
%
We assume that
\begin{enumerate}
    \item  $T_{\text{max}} \approx 1$ in an optimal parameter regime, as motivated by the results of Figure \ref{fig:TvskAcceleratedRDB}; 
    \item  The optimum $\tilde{F}_C$ corresponds to the position of the resonant transmission peak $T_{\text{max}} \approx 1$, as shown in Figure \ref{fig:FcTvsk}a). From equation \eqref{eq:Transmissioncoefficientk}, this approximation corresponds to $\Phi(a) = n\pi$ for $n \in \mathbb{Z}$.
\end{enumerate}
In the limit $\delta a_0 \rightarrow 0$, we obtain 
\begin{align}
    \tilde{F}_{C_{\text{opt}}}&=\frac{4\tilde{d}^2\mathcal{F}^2 \tilde{L}^2 }{(\tilde{k}_i^2+\tilde{a}_0\tilde{L})\pi^2}.
    \label{eq:optimalFCDimensionless}
\end{align}
Converting back to dimensional form gives
\begin{align}
	F_{C_{\text{opt}}}&=\frac{16m^4d^2\mathcal{F}^2 L^2}{\hbar^4 \pi^2 \left[ {k_i}^2 + \frac{2m^2 L a_0}{\hbar^2} \right]}.	 \label{eq:approxoptimalFC}
\end{align}
This is a key result of this paper, and can be used to efficiently determine the fundamental acceleration sensitivity attainable by an atomic FPI.

\subsection{Comparing Acceleration Sensitivities of an Atomic FPI and a MZ Interferometer}
\label{Sec32}
\begin{figure}[t!]
	\centering
\includegraphics[width=0.8\linewidth]{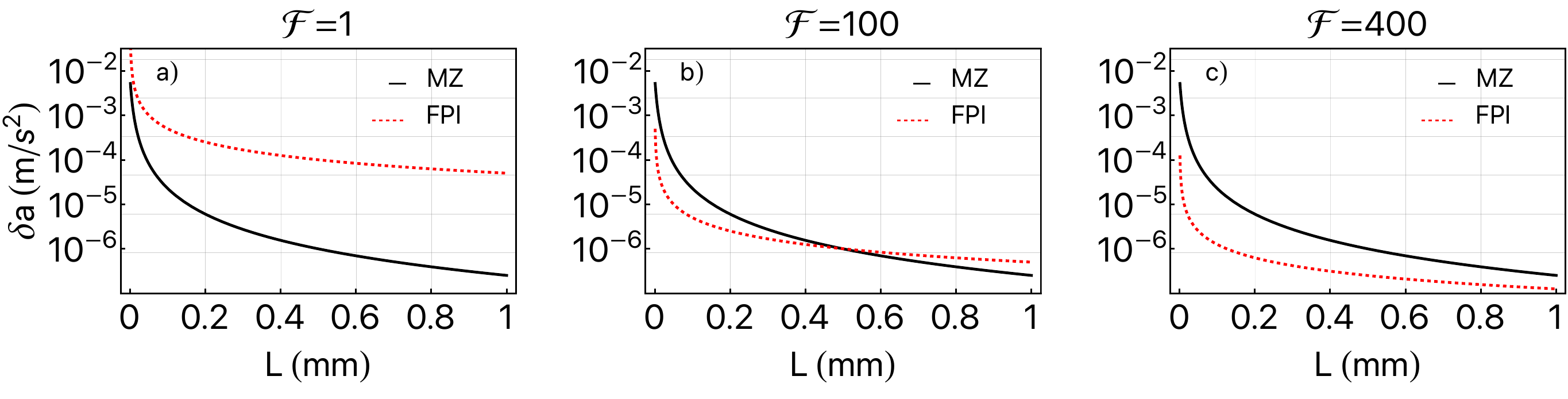}
\caption[Comparison of the  acceleration sensitivities of a space-based Mach-Zehnder interferometer and an atomic FPI]{Plots comparing the acceleration sensitivities of a space-based ($a_0\rightarrow 0$) MZ interferometer (black curve) and an atomic FPI (red dashed curve) with $k_i=0.0596 k_0$, $w = 1\upmu \text{m}$ and $d = 4\upmu \text{m}$. Here the minimum change in acceleration that can be detected, $\delta a$, is plotted as a function of length $L$, for varying finesse values a) $\mathcal{F}=1$, b) $\mathcal{F}=100$ and c) $\mathcal{F}=400$. $L$ is the initial distance between the BEC and the first barrier in an atomic FPI. For a MZ interferometer, the total interferometer length $ L_t=L+2w+d $. A particular finesse can be achieved while varying $L$, by manipulating barrier height accordingly. The sensitivity of an atomic FPI is better than that of the comparison MZ interferometer for high finesse and low length scales.}
\label{fig:deltaaMzFPIa0}
\end{figure}

\label{sec:zerowidthcomparison}
Having obtained an expression for the optimal Fisher information for atomic FPI, we may now compare that result with the optimum sensitivity of a MZ interferometer of equivalent interferometer length, since total device length is the limiting constraint for many real-world applications. For concreteness, we consider space-based accelerometry, which is a particularly tight-SWaP deployment environment.

For a space-based system, we can approximate $a_0 = 0$, and hence the $F_{C_{\text{opt}}}$ and optimum sensitivity per particle of an atomic FPI can be found from equations \eqref{eq:deltaa} and \eqref{eq:approxoptimalFC}: 
\begin{subequations}
\begin{align}
	F_{C_{\text{opt}}}&=\frac{16m^4d^2\mathcal{F}^2 L^2}{\hbar^4 \pi^2 k_i^2},\\
	\delta a_{\text{FPI}} &=\frac{1}{\sqrt{F_{C_{\text{opt}}}}} =\frac{k_i \pi \hbar^2}{4m^2 d\mathcal{F}L}.	 \label{Eq.deltaaFPIa00}
\end{align}
\end{subequations}
In the case of a MZ interferometer, the per particle sensitivity is given by \cite{peters_high-precision_2001} 
\begin{align}
	\delta a_{\text{MZ}}&=\frac{1}{k_i \mathbb{T}^2} \label{Eq.deltaaMZ},
\end{align}
where $k_i$ is the effective momentum transferred to the atoms by the beamsplitters of the interferometer. The total interferometer time, $2\mathbb{T}$ is constrained to
\begin{align}
	\mathbb{T}&=\frac{L_\text{t}}{v},
\end{align}
where $v=\hbar k_i/m$ is the velocity of the atoms when $a_0 \rightarrow 0$, and $L_t$ is the total length of the interferometer. For a fair comparison, we set $L_t$ to the same total length as the atomic FPI: $L_\text{t}=L+2w+d$. Substituting these into equation \eqref{Eq.deltaaMZ} gives
\begin{align}
	\delta a_{\text{MZ}} &= \frac{\hbar^2 k_i}{m^2(L+2w+d)^2} \label{Eq.deltaaMZ2}.
\end{align}
We can see immediately that making the device size $L_t$ smaller results in poorer sensitivity for both a MZ interferometer and an atomic FPI. However, equations \eqref{Eq.deltaaFPIa00} and \eqref{Eq.deltaaMZ2} show that the scaling of sensitivity with device length is different in each case. We are therefore motivated to study the variation of sensitivity $\delta a$ with length $L$, which we show for a range of finesse values $\mathcal{F}$ in Figure \ref{fig:deltaaMzFPIa0}. Here optimum values are used for $k_i, w$ and $d$ \cite{manju_atomic_2020}. For low length ($L < 0.4 \text{mm}$) and high finesse ($\mathcal{F} > 100$), the fundamental sensitivity of an atomic FPI surpasses that of a MZ interferometer.

\subsection{Space-Based Accelerometer Using a BEC With a Finite Momentum Width}
\label{sec:finitemomentumwidth}
Having compared an atomic FPI to a MZ interferometer for the ideal case of infinitely narrow momentum width, we now consider the more realistic case of finite momentum width. This immediately presents a new challenge: if this finite momentum width is greater than the cavity linewidth, only a portion of the atomic cloud (in momentum space) is on resonance and gets transmitted. This reduces the resonant transmission (i.e. $T_{\text{max}} < 1$), so in order to achieve complete resonant transmission (i.e. $T_{\text{max}} = 1$) we are immediately restricted to cavities with linewidth greater than the finite momentum width of the cloud. As linewidth decreases as cavity width increases, this puts an upper bound on the cavity width for an effective atomic FPI \cite{manju_atomic_2020}. For the cloud of finite momentum width, we consider a Gaussian of full width at half maximum (FWHM) of $\Delta k = 0.008k_0$, where here $k_0 = 2\pi/(780\text{nm})$. Note that $\Delta k = 2 \sqrt{2 \ln 2}/\sigma_c$, where $\sigma_c/\sqrt{2}$ is the standard deviation of the Gaussian density profile, or equivalently a Gaussian $k$-space density profile with standard deviation $1/(\sqrt{2}\sigma_c)$. \\

With that in mind, we now study how the atomic cloud's finite momentum width affects the dependence of the optimal Fisher information $\tilde{F}_C$ on the initial momentum kick $\tilde{k}_i$ and cavity width $\tilde{d}$. 
\begin{figure}[t!]
	\centering
 	\includegraphics[width=0.8\linewidth]{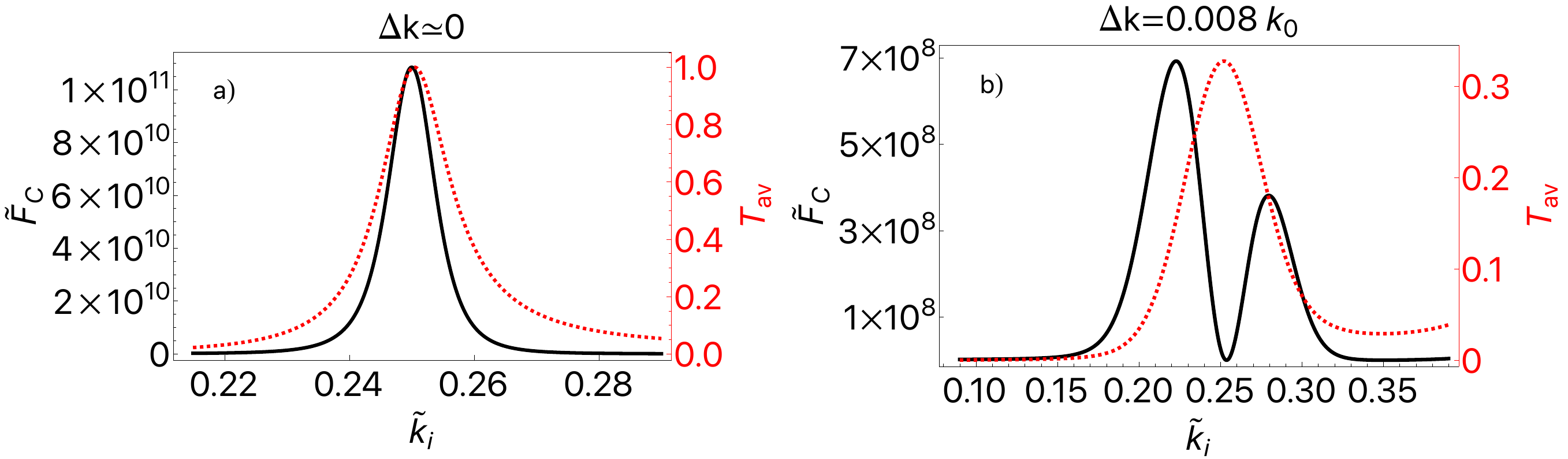}

	\caption[Position of the optimum $F_C$ with respect to the transmission resonance peak]{Transmission coefficient (red dashed curve) and classical Fisher information (black curve) are plotted as a function of the atomic wave vector, a) for an infinitely narrow momentum width atomic cloud and b) for a finite momentum width atomic cloud. The optimum $\tilde{F}_C$ is positioned almost at the $\tilde{k}_i$ corresponding to the transmission peak in the first case. However, it deviates from the transmission peak point when the cloud has a finite momentum width. Here, $k_0=2\pi/(780 \text{nm})$.}
	\label{fig:FcTvsk}
\end{figure}
In Figure~\ref{fig:FcTvsk}, we plot $\tilde{F}_C$ as a function $\tilde{k}_i$ for an atomic cloud with infinitely narrow and finite momentum width in Figure \ref{fig:FcTvsk}a) and Figure \ref{fig:FcTvsk}b) respectively. We use the parameters $w \kappa = 1$, $w = 1 \upmu \text{m}$ and $\kappa = 0.124 k_0$, as determined in Ref. \cite{manju_atomic_2020}. \\

In Figure \ref{fig:FcTvsk}b), we demonstrate that for a finite momentum width source, the location of the optimum Fisher information $\tilde{F}_C$ shifts to no longer coincide with the location of the transmission peak. $\tilde{F}_C$ is now relatively small at the position of $T_{\text{max}}$, and the optimum $\tilde{F}_C$ is shifted to a point where the slope of the resonance curve is non-zero. To explain this shift, recall equation \eqref{eq:FC}: 
\begin{align}
    F_C(a) &=\frac{(\partial T/\partial a)^2}{T(a)}+\frac{(\partial T/\partial a)^2}{1-T(a)}. 
\end{align}
Around the transmission peak point, the curve can be approximated as a quadratic, 
\begin{align}
T(a)&=T_{\text{max}}(1-Ca^2),
\end{align}
where $C$ is a constant .Under a uniform acceleration, $k$ and $a$ have a roughly linear relationship, so at the transmission peak $T_{\text{max}}$ the derivative $\partial T / \partial a \propto \partial T/ \partial k \approx 0$. For an infinitely narrow momentum width source, $T_{\text{max}} = 1$, so the denominator in the Fisher information $T(1-T) = 0$. Then, 
\begin{align}
    \lim_{a\to 0} F_C &= \frac{0}{1}+\lim_{a\to 0}\frac{(-2Ca)^2}{Ca^2}=4C.
\end{align}
Hence, $F_C$ converges to a finite value. However, a source with non-zero momentum width has $T_{\text{max}} < 1$, so only the numerator approaches zero and $F_C \rightarrow 0$ at the transmission peak. Thus, the peak Fisher information does not coincide with the transmission peak; so for a finite momentum width source, we have to numerically determine the parameter regime which optimises $F_C$.
\begin{figure}[t!]
	\centering
	\includegraphics[width=0.8\linewidth]{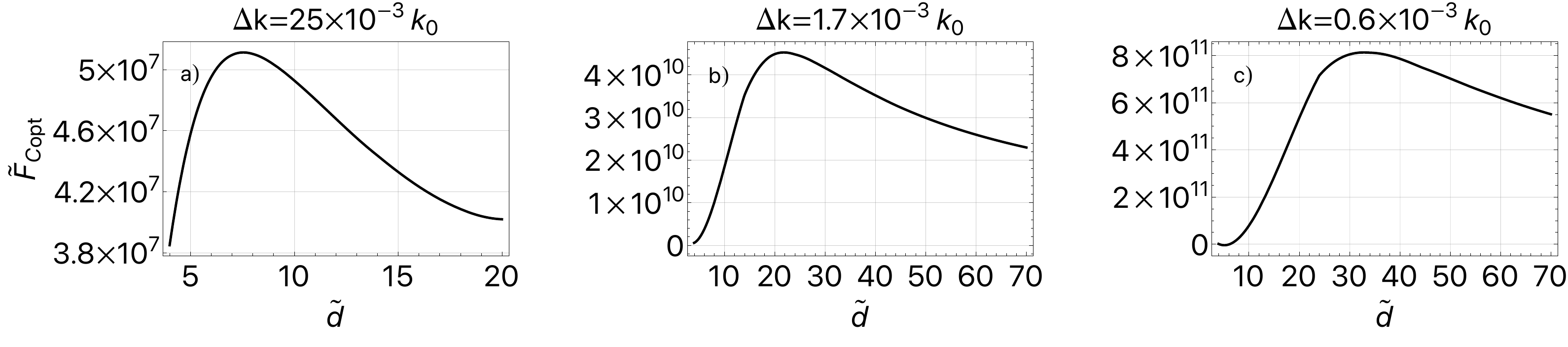}
	\caption[ The optimum $\tilde{F}_C$ as a function of the cavity length, for different momentum widths]{Optimum value of $\tilde{F}_C$ as a function of cavity length, for different momentum widths. $\tilde{F}_{C_{\text{opt}}}$ increases up to a maximum value and then decreases with increasing cavity length. The optimum cavity length which gives $\tilde{F}_{C_{\text{opt}}}$ varies with the momentum width. Here, $k_0=2\pi/(780\text{nm})$.}
	\label{fig:FcOptvsdFiniteSigma}
\end{figure} 
\subsection{Optimising Cavity Length and Corresponding Acceleration Sensitivity for a Particular Momentum Width}
\label{sec:Optdfordeltak}
As before, we calculate the Fisher information $\tilde{F}_C$ as a function of initial momentum kick $\tilde{k}_i$ for a range of cavity widths $\tilde{d}$ and momentum widths $\Delta k = 2 \sqrt{2 \ln 2}/\sigma_c$ which gives a maximum $\tilde{F}_C$. For each cavity length there is an optimum $\tilde{k}_i$ corresponding to the optimum Fisher information $\tilde{F}_C$ and this optimum point changes with cavity length. However, the optimum $\tilde{k}_i$ is not changed by variation in the momentum width $\Delta k$. This is because the optimum initial momentum kick $\tilde{k}_i$ depends upon the location of the first resonant peak, which is independent of the momentum width of the source. However, the momentum width $\Delta k$ does affect the height of the transmission peak, and therefore the optimum $\tilde{F}_C$ \cite{manju_atomic_2020}.
We thus plot the maximum $\tilde{F}_C$ that occurs at the optimum $\tilde{k}_i$ as a function of cavity width $\tilde{d}$ in Figure \ref{fig:FcOptvsdFiniteSigma} for different source momentum widths $\Delta k$.\\

Unlike in the infinitely narrow momentum width case (see Figure \ref{fig:Optk0Fcvsd}a)), there is not an unbounded increase in the optimum Fisher information $\tilde{F}_C$ with cavity width $\tilde{d}$. Rather, $\tilde{F}_C$ reaches a maximum and then decreases after the cavity width $\tilde{d}$ exceeds an optimal value. This optimal value arises because the cavity linewidth decreases as the cavity width increases \cite{manju_atomic_2020}. When the cavity width $\tilde{d}$ exceeds the optimal value, the corresponding linewidth becomes smaller than the momentum width $\Delta k$, and (as discussed earlier) transmission through the atomic FPI is suppressed and the optimal Fisher information $\tilde{F}_{C_{\text{opt}}}$ decreases. \\

\begin{figure}[t!]
	\centering
	\includegraphics[width=0.8\linewidth]{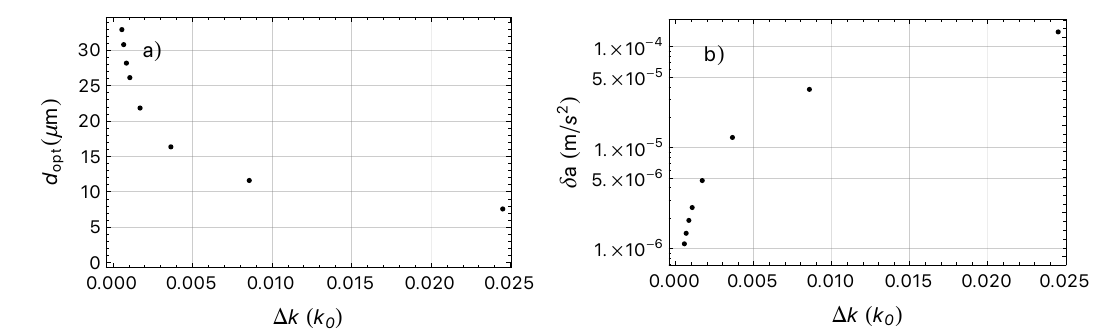}
	\caption[The optimum cavity length and acceleration sensitivity as a function of the BEC momentum width]{a) Plot showing the optimum value of the cavity length (which gives maximum $F_C$) as a function of the momentum width of the BEC source. The optimum cavity length decreases with increasing momentum width. b) Minimum detectable change in acceleration, $\delta a$ decreases with decreasing momentum width of the BEC source. Hence acceleration sensitivity improves with decreasing momentum width. Here, $k_0=2\pi/(780 \text{nm})$.}
	\label{fig:doptdeltaavsdeltak}
\end{figure}

Equivalently, as the momentum width increases, a larger linewidth (and smaller cavity width) is needed for complete transmission. That is, the peak in Fisher information $\tilde{F}_C$ occurs at a smaller cavity width. This trend is illustrated in Figure \ref{fig:doptdeltaavsdeltak}a), where we plot the cavity width $d$ corresponding to the maximum optimal Fisher information against the momentum width $\Delta k$. \\

We also find in Figure \ref{fig:FcOptvsdFiniteSigma} that the maximum optimal Fisher information  decreases as the momentum width increases. This is because, as shown in Ref. \cite{manju_atomic_2020}, resonant transmission is suppressed and transmission peaks broaden as momentum width $\Delta k$ increases. A decrease in Fisher information corresponds to a worse sensitivity (i.e. a larger $\delta a$), as illustrated in Figure~\ref{fig:doptdeltaavsdeltak} by plotting $\delta a$ as a function of momentum width $\Delta k$. 
\\\\
\subsection{Comparing Sensitivities of a MZ Interferometer and an Atomic FPI}
\label{sec:finitewidthcomparison}
Having determined the optimum cavity width $d$ for each momentum width $\Delta k$, the device length $L_t$ approximately given by the parameter $L$. We can now study the relationship between optimum Fisher information and length $L$.
\begin{figure}
	\centering
	\includegraphics[width=1\linewidth]{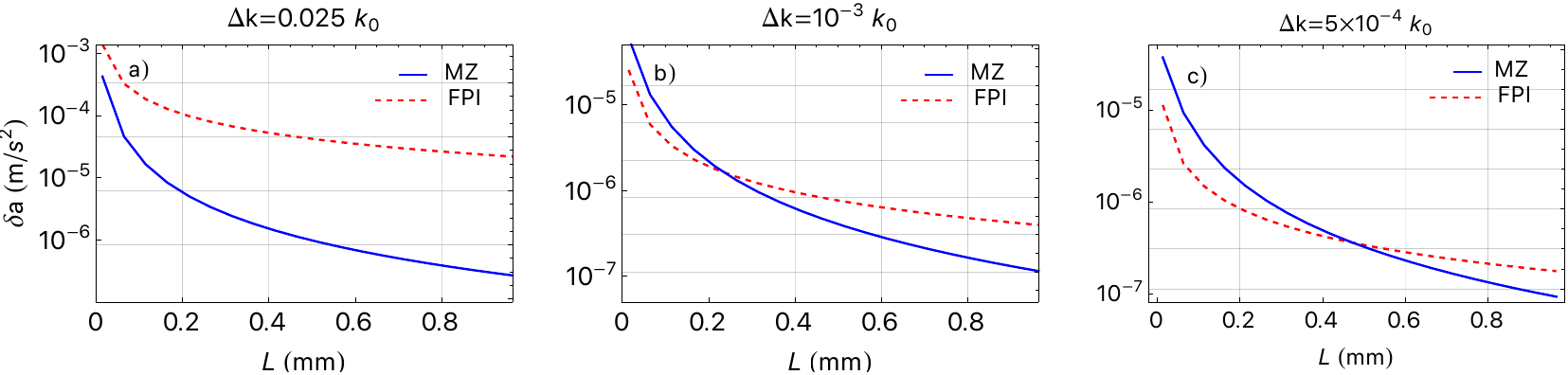}
	\caption[Acceleration sensitivities of a Mach-Zehnder interferometer vs an atomic FPI using a finite momentum width BEC]{Plot comparing the acceleration sensitivity as a function of length $L$ of an atomic FPI with that of a MZ interferometer, using a finite momentum width BEC source for a space based system ($a_0 \rightarrow 0$). a), b) and c) use  $d \kappa=\{7.55, \,26.1, \,32.89\}$ and $k_i/\kappa=\{0.237,\, 0.106,\, 0.086\}$ respectively, which are the optimum values corresponding their specific momentum widths. Three plots compare the sensitivities using BECs with different momentum widths. For small device length and low momentum width, the sensitivity of an atomic FPI surpasses that of a MZ interferometer.}
	\label{fig:deltaaMZFPIdeltak}
\end{figure} 
In Figure \ref{fig:deltaaMZFPIdeltak}, we compare the sensitivity of an atomic FPI to a MZ interferometer of equivalent length $L$ for three different momentum widths $\Delta k$. For a currently experimentally-achievable momentum width ($\Delta k = 0.025 k_0$), the MZ device achieves superior sensitivity. However, for an order of magnitude narrower momentum width ($\Delta k = 10^{-3} k_0$) an atomic FPI can compete with a MZ device for small lengths, and for an order of magnitude further ($\Delta k = 5 \times 10^{-4}k_0$) the atomic FPI is more sensitive for $L < 0.2\text{mm}$. Thus, the atomic FPI presents a future alternative for more sensitive space-based accelerometry, once cooling schemes can further narrow the momentum width of atomic clouds.

\section{Conclusions}
We have investigated the application of an atomic FPI as a space-based acceleration sensor with a non-interacting pulsed BEC source. We used an analytic approximation for the transmission of the FPI, and showed this expression agreed well with the exact transmission predicted by simulation of the Schr\"{o}dinger equation. We quantified the sensitivity of the atomic FPI as an accelerometer using the classical Fisher information, and studied the sensitivity of FPIs with an atomic cloud source of both infinitely narrow and finite momentum width. In the case of an infinitely narrow momentum width source, an approximate expression for the optimum Fisher information and sensitivity was derived, and it was found that in the low length and high finesse regime an atomic FPI provided superior sensitivity to a MZ interferometer of equivalent total device length.

When a finite momentum width source is used, there is suppression of the transmission, particularly when the momentum width exceeds the linewidth of the cavity, leading to an optimal cavity width for maximum sensitivity. For a presently achievable momentum width, the atomic FPI cannot beat the sensitivity of a MZ interferometer under optimal parameter regimes. However, if a narrower momentum width can be realised, the atomic FPI could achieve greater sensitivity in the low length regime. Therefore, if atomic cooling techniques continue to advance in the coming years, the atomic FPI presents an exciting alternative as an accelerometer with a superior sensitivity to a MZ device.

These results are a best-case scenario: we have assumed a non-interacting BEC, and have only worked in 1D. Detailed analysis of the uncertainty in the cavity length, barrier height, and velocity of the atomic cloud are necessary in future work.

This work has focused on the atomic FPI as a space-based accelerometer. When applied as an earth-based accelerometer, the atomic cloud gains significantly more energy in the accelerating period before the first barrier. The barrier height must then be increased to exceed the atomic energy (a necessary condition for the atomic FPI to function), which necessitates a reduction in barrier width and cavity length as per the parameters determined in Ref.~\cite{manju_atomic_2020}. The barrier width required now is three orders of magnitude smaller than achievable by existing laser focusing methods. To overcome this difficulty, we propose an alternative scheme: initially position the atoms on the right hand side of the barrier in Figure \ref{fig:SchematicFPIAccelerating}, and provide a momentum kick against the direction of the acceleration (towards the barriers). The acceleration will therefore reduce the energy of the atomic cloud, so that an experimentally feasible barrier width may be used.

\begin{acknowledgments}
SAH acknowledges support through an Australian Research Council Future Fellowship Grant No. FT210100809. SSS was supported by an Australian Research Council Discovery Early Career Researcher Award (DECRA), Project No. DE200100495. This research was undertaken with the assistance of resources and services from the National Computational Infrastructure (NCI), which is supported by the Australian Government.
\end{acknowledgments}

\bibliography{bib.bib}

\end{document}